\documentclass[12pt]{article}
\title{Enhancement of zonal flow damping due to resonant magnetic perturbations in the background of an equilibrium $E \times B$ sheared flow}

\author{M. Leconte and R. Singh\\
National Fusion Research Institute, Daejeon 34133, South Korea \\
\textit{Email: mleconte@nfri.re.kr}}

\usepackage{graphicx}
\usepackage{amsmath}

\newcommand{\dif}{\partial}
\newcommand{\psih}{\psi_h}
\newcommand{\phih}{\phi_h}

\newcommand{\ky}{k_0}
\newcommand{\qx}{q_x}
\newcommand{\ve}{V_E}
\newcommand{\heta}{\eta}
\newcommand{\hnu}{\nu}

\newcommand{\zp}{\phi_q}
\newcommand{\freqzf}{\Omega}

\renewcommand{\ln}{L_n}

\renewcommand{\wp}{\omega_1}
\newcommand{\wm}{\omega_2}
\newcommand{\kp}{k_1}
\newcommand{\km}{k_2}

\newcommand{\amp}{\Psi}
\newcommand{\pot}{\Phi}
\newcommand{\freq}{\delta}

\newcommand{\hmu}{\mu}

\newcommand{\hnl}{\Lambda}

\newcommand{\aq}{\alpha_{DW}}

\newcommand{\fq}{\phi_q}

\begin{document}
\maketitle

\begin{abstract}
Using  a parametric interaction formalism, we show that the equilibrium sheared rotation can enhance the zonal flow damping effect found in Ref. [M. Leconte and P.H. Diamond, \emph{Phys. Plasmas} 19, 055903 (2012)]. This additional damping contribution is proportional to $(L_s/L_V)^2 \times \delta B_r^2 / B^2$, where $L_s/L_V$ is the ratio of magnetic shear length to the scale-length of equilibrium $E \times B$ flow shear, and $\delta B_r / B$ is the amplitude of the external magnetic perturbation normalized to the background magnetic field.
\end{abstract}

\section{Introduction}

The high confinement mode (H-mode) regime is a reference operation scenario for future tokamak experiments like ITER. In this regime, boundary disturbances known as Edge Localized Modes need to be either avoided or controlled. One candidate control method uses external magnetic perturbations known as Resonant Magnetic Perturbations i.e. RMP \cite{EvansRev2015}. RMPs were shown to damp GAM zonal flows \cite{Xu2011,Robinson2012} and to enhance turbulent density fluctuations \cite{McKee2013,Wilcox2018,Schmitz2018}.
The modification of turbulence and flows was also observed for the case of a large-scale static magnetic island $m:n = 2:1$ \cite{Zhao2015, Choi2017, Kwon2018}.
Proposed mechanisms to explain the enhanced zonal flow damping include modifications to the Rosenbluth-Hinton residual zonal flows due to 3D magnetic geometry \cite{SugamaWatanabe2006,ChoiHahm2018,Terry2013}.
An alternative mechanism was described by Leconte \& Diamond \cite{LeconteDiamond2012}.
In this article, we recast the latter theory in the language of parametric interaction, and consider important additional effects on zonal flows due to the synergy between RMPs and the equilibrium flow shear.
Our proposed physical mechanism for RMP-induced zonal flow damping by a large-scale static magnetic perturbation can be understood as the \emph{coupling} of the zonal flow branch to a damped Alfven wave branch. In the parametric interaction analysis, the sideband response to zonal potential $\phi_q$ scales as $\psi_s \sim \psi_0^* \phi_q / B \times (\Omega +i \eta q_r^2)^{-1}$ with $\psi_0$ the external magnetic perturbation, $q_r$ the radial wavenumber of zonal flows, $\Omega = \Omega_r + i \gamma_q$ the complex frequency of the modulational instability, and $\eta$ the Spitzer resistivity. Replacing $\psi_s$ in the expression for the Maxwell stress , the resulting dispersion relation is: $(\Omega-i \gamma_q^0)(\Omega+i \eta q_r^2) = c_A^2 |\tilde B_r|^2 / B^2$, where $\gamma_q^0$ denotes the unperturbed zonal flow growth-rate (balance between turbulence drive and neoclassical damping), and $c_A = B/\sqrt{4 \pi n_0 m_i}$ is the Alfven speed. Without magnetic perturbation $|\tilde B_r| = 0$, the two branches decouple, one becomes the usual turbulence-driven zonal flow branch $\Omega \sim i \gamma_q^0$, while the other is purely damped $\Omega \sim -i \eta q_r^2$.
This shows that the external magnetic perturbation effectively couples the two branches of the dispersion relation, resulting in a modification of zonal flow growth.

\section{Model}

In order to calculate the effect of the nonlinear Reynolds stress due to 3D fields on zonal flow dynamics, we first present the linear relation between an externally-imposed $\delta \psi_0$ perturbation and a helical stream function $\delta \phih$ in weak shear rotating plasmas. We consider a slab geometry ($x,y,z$), where $x$ denotes the local radial coordinate, and $y$ denotes the local poloidal coordinate, and $z$ is the local toroidal coordinate, in a fusion device.
We introduce a flux function $\psi$, via ${\bf B}_\perp = - \hat {\bf z} \times \nabla \psi$ and a stream-function $\phi$, with ${\bf v}_E = \hat {\bf z} \times \nabla \phi$. For small perturbations, the flux and stream functions are expressed as $\psi = \psi_{eq}+ \delta \psih$ and $\phi = \phi_{eq} + \delta \phih$, with the mean magnetic flux $\psi_{eq}(x) = B_0 x^2 / 2 L_s$.
Here, we impose the radial profile of the mean stream function as $(c/B) \phi_{eq}(x) = \frac{V_{E0}}{L_V} [\frac{x^2}{2} +\frac{x^3}{6L_V} + \ldots]$ ($E \times B$ sheared flow) near the rational surface \cite{HuWangWei2016, Singh2018}. For weak flow shear $\Delta x / L_V <1$, with $\Delta x$ the characteristic radial scale, and in the limit of low-resistivity, like the frozen-in dynamics, i.e. $E_\parallel=0$, the linearized Ohm's law is given by:
$\hat {\bf z} \times \nabla \phi_{eq} \cdot \nabla \delta \psi_h + \hat {\bf z} \times \nabla \delta \phih \cdot \nabla \psi_{eq} = 0$.
The $E \times B$ flow is related to $\phi_{eq}$ via $V_E(x)= (c/B) \phi'_{eq}(x) \simeq V'_E x$ with $V'_E = V_0 x / L_V$.
Integrating between parallel and poloidal coordinates, this equation then gives the linear relation:
$\delta \phih = \frac{L_s V_{E0}}{L_V} \delta \psi_h$.

Using normalizations $\phih = (e \delta \phi_h / T_e) (\ln / \rho_s)$ and $\psi_h = (e \delta \psi_h / T_e)(\ln / \rho_s)(2c_s / c \beta)$, with $\ln=n_0 / |\nabla n_0|$ the density-gradient length, we obtain the desired relation - in normalized form - which is later used in the zonal flow dynamics.
\begin{equation}
\phih = \frac{\beta}{2} |V'| \psih.
\label{ident1}
\end{equation}
with the normalized flow-shear $V' = (L_s / L_V) V_{E0} / c_s$, $L_s = q_0 R$. Note that Eq. (\ref{ident1}) is valid for $\Delta x \sim \Delta w < L_V$ (small island width $\Delta w$), with $\ln=n_0 / |\nabla n_0|$ the density-gradient length.


Writing the magnetic field ${\bf B} = {\bf B}_0 - \hat {\bf z} \times \nabla \psi$, and the current $j_\parallel= \nabla_\perp^2 \psi$, where $\psi$ is the magnetic flux and ${\bf B}_0 = B_0 \hat {\bf z}$ is the toroidal magnetic field, the model equations for coupled zonal flows and external magnetic perturbation are the vorticity equation (charge balance) and Ohm's law:
\begin{eqnarray}
\frac{\dif \nabla_\perp^2 \tilde \phi_k }{\dif t} - \nabla_{\parallel0} \nabla_\perp^2 \tilde \psi_k =
\frac{\beta}{2} \{ \tilde \psi, \nabla_\perp^2 \tilde \psi \}  - \{ \tilde \phi, \nabla_\perp^2 \tilde \phi \} + \hnu \nabla_\perp^4 \phi_k,
\label{cb0} \\
\frac{\dif \psi_k}{\dif t} - \frac{2}{\beta} \nabla_{\parallel0} \phi_k =  - \{ \tilde \phi, \tilde \psi \} + \heta j_{\parallel k}.
\label{ohm0}
\end{eqnarray}
Here, we have written the perturbed parallel gradient as $\nabla_\parallel = \nabla_{||0} + \frac{ {\bf \delta B} }{B} \cdot \nabla$, with $\nabla_{||0}$ the equilibrium part, and $\frac{ {\bf \delta B} }{B} \cdot \nabla$ the contribution due to external magnetic perturbations. Time is normalized as $\frac{c_s}{\ln} t \to t$, and the perpendicular and parallel scales are normalized respectively as: $\rho_s \nabla_\perp \to \nabla_\perp$ and $L_n \nabla_{\parallel0} \to \nabla_{\parallel0}$.
Other normalizations are:
$\frac{L_n}{\rho_s} \frac{e}{T_e} \phi \to \phi$, $\frac{2 c_s L_n}{c \rho_s \beta_e} \frac{e}{T_e} \psi \to \psi$, and $\frac{\ln}{\rho_s c_s} V_E \to \ve$, with $\displaystyle \beta = \frac{8 \pi n_0 T_e}{B^2}$ the ratio of kinetic to magnetic energy, $\displaystyle \hat \heta = \frac{\eta c^2 \ln}{4 \pi \rho_s^2 c_s} = \frac{\lambda_s^2}{\rho_s^2} \frac{\nu_{ei} \ln}{c_s}$ the normalized resistivity,
, $\lambda_s = \frac{c}{\omega_{pe}}$ the electron skin-depth
and $\hat \nu= \frac{\ln}{\rho_s^2 c_s} \nu$ the normalized - turbulent - viscosity. 
In the following, we drop the $\hat{}$ on normalized quantities for clarity.

\begin{figure}
\begin{center}
\begin{tabular}{cc}
\includegraphics[width=0.5\linewidth]{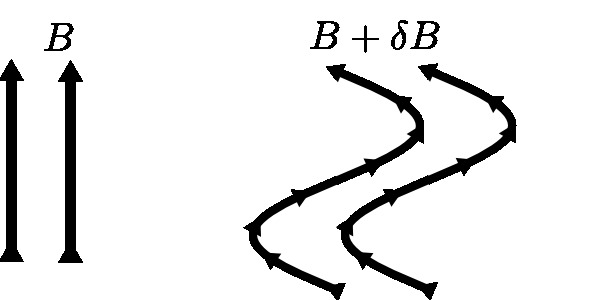}
\end{tabular}
\caption{The radial perturbation $\delta B_r$ due to 3D fields can be viewed as a long-wavelength modulation of the background magnetic field $\bf B$.}
\label{sketch}
\end{center}
\end{figure}

To describe the perturbed flux surface geometry due to RMPs, we use the following ansatz for the total magnetic flux:
\begin{equation}
\psi = \psi_{eq} + \psih \cos \ky y
\end{equation}
where $\psi_{eq}$ is the unperturbed poloidal flux, and $\ky$ is the poloidal wavenumber of the RMP.
Physically, this represents a long-wavelength modulation of the magnetic field [Fig. \ref{sketch}].

\subsection{Parametric interaction analysis}

\begin{figure}
\begin{center}
\begin{tabular}{cc}
\includegraphics[width=0.5\linewidth]{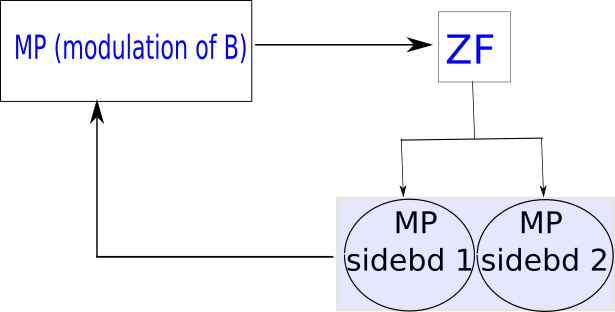}
\end{tabular}
\caption{Schematic diagram of the parametric interaction.}
\label{diagram}
\end{center}
\end{figure}

We vizualize the interaction between zonal flows and external magnetic perturbation as a four-wave parametric interaction \cite{LashmoreDavies2001, Singh2017, Singh2001}. A schematic diagram of the interaction is shown [Fig. \ref{diagram}].
Parametric interaction is associated with a phase-instability, as described e.g. in Ref. \cite{CrossGreenside2009}.
Here, the magnetic perturbation acts like a stationary long-wavelength modulation of the background magnetic field $(\omega_0=0, {\bf k}_0=k_0 \hat {\bf y}  )$, and zonal flows act like a long-scale wave $(\Omega, {\bf q} = q_x \hat {\bf x})$. Here, $k_0 \gg q_y$, since $q_y=0$ for zonal flows. In  practical experiments, we expect the magnetic perturbation to evolve in time, but since this evolution is very slow, we can treat it as stationary. The MP can be called the pump, although it provides a damping rather than a drive, as far as zonal flows are concerned.
In this parametric process, a long-scale wave (ZF) at $(\freqzf, {\bf q})$ interacts with the magnetic perturbation at $(\omega_0=0, {\bf k}_0 )$ and generates two side-bands at $(\wp,\kp)$ and $(\wm,\km)$, where ${\bf k}_{1,2} = {\bf q} \pm {\bf k}_0$ and $\omega_{1,2} = \omega({\bf k}_{1,2})$. The resonant interaction condition for the waves $\omega_0 \pm {\rm Re} \{ \Omega \} = {\rm Re} \{ \omega_{1,2} \}$ is approximately satisfied since, to first approximation, zonal flows have zero frequency.
The two side-bands couple with the magnetic perturbation to produce electrostatic and magnetostatic ponderomotive forces (poloidal torques) on the plasma, which can excite and/or damp the low-frequency mode (ZF).  

Let us now obtain the equations for the two sideband amplitudes, following the notations of Lashmore-Davies \emph{et al.} \cite{LashmoreDavies2001}.
To calculate the parametric interaction between short-scale MP and long-scale zonal flows, we take the magnetic perturbation as:
\begin{eqnarray}
\tilde \psih & = & \psih(x) [ \exp(i \ky y-i \omega_0 t) +c.c. ]
\label{psipump} \\
\tilde \phih & = & \phih(x) [ \exp(i \ky y-i \omega_0 t) +c.c. ] 
\label{phipump}
\end{eqnarray}
Here, the pump frequency is $\omega_0 \sim 0$, for static MPs. This mode can couple to long scale zonal wave, which is represented by:
\begin{eqnarray}
V_{ZF}(x,t) = i \qx\fq(t) \exp(i \qx x-i \Omega t) +c.c.
\label{zfdef} \
\end{eqnarray}
where $\qx$ is the dimensionless wavenumber along the radial direction, and $\phi_q$ is the potential amplitude.

The resonant coupling between the pump mode ($\omega_0, {\bf k}_0$) and zonal flow ($\Omega, {\bf q}$) can generate two sideband waves: $(\omega_{1,2}), {\bf k}_{1,2}$, $\omega_{1,2}= \omega_0 \pm \Omega$, with ${\bf k}_{1,2} = {\bf q} \pm {\bf k}_0$. The desired sideband field $(\psi_{1,2}, \phi_{1,2})$ can be represented as:
\begin{eqnarray}
\tilde \psi_{1,2}({\bf r},t)  = \psi_{1,2} [ \exp (i {\bf k}_{1,2} \cdot {\bf r} - i \omega_{1,2} t) + c.c.]
\label{sbdef1} \\
\tilde \phi_{1,2}({\bf r},t)  = \phi_{1,2} [ \exp (i {\bf k}_{1,2} \cdot {\bf r} - i \omega_{1,2} t) +c.c.]
\label{sbdef2}
\end{eqnarray}

Using equations (\ref{psipump},\ref{phipump},\ref{zfdef},\ref{sbdef1},\ref{sbdef2}), the vorticity equation (\ref{cb0}) for zonal flow can be written:
\begin{equation}
\frac{\dif \fq}{\dif t} = - \frac{\beta}{4} (\hat {\bf z} \times {\bf q}) \cdot {\bf k}_0 (\psih^* \psi_1 - \psih \psi_2)
+\frac{1}{2} (\hat {\bf z} \times {\bf q}) \cdot {\bf k}_0 (\phih^* \phi_1 - \phih \phi_2)
\end{equation}

Note that the subscript "h" corresponds to the externally applied helical perturbation. The perturbation with subscript "q" represents the zonal flow, and the field with subscript "1,2" represents the driven sideband perturbation.

From Ohm's law Eq. (\ref{ohm0}), the equations of the two sideband waves $\psi_1, \phi_1$ and $\psi_2, \phi_2$ are:
\begin{eqnarray}
\frac{\dif \psi_1}{\dif t} + \eta k_1^2\psi_1 & = & (\hat {\bf z} \times {\bf q}) \cdot {\bf k}_0 ~\psih \fq \\
\frac{\dif \psi_2}{\dif t} + \eta k_2^2\psi_2 & = & - (\hat {\bf z} \times {\bf q}) \cdot {\bf k}_0 ~ \psih^* \fq \\
\frac{\dif \phi_1}{\dif t} + \hnu \kp^2 \phi_1 & = & (\hat {\bf z} \times {\bf q}) \cdot {\bf k}_0 \frac{\ky^2-q^2}{k_1^2} \phih \fq \\
\frac{\dif \phi_2}{\dif t} + \hnu \km^2 \phi_2 & = & - (\hat {\bf z} \times {\bf q}) \cdot {\bf k}_0 \frac{\ky^2-q^2}{k_2^2} \phih^* \fq,
\end{eqnarray}

where the sideband complex amplitudes $\psi_{1,2}$ can be further decomposed as:
\begin{equation}
\begin{bmatrix}
\psi_{1,2}(t) \\
\phi_{1,2}(t)
\end{bmatrix}
=
\begin{bmatrix}
\amp_{1,2}(t) \\
\pot_{1,2}(t)
\end{bmatrix}
e^{i \freq_{1,2} t},
\end{equation}

with $\freq_{1,2} = \omega_{1,2} - \omega_0$,
and the unperturbed frequencies $\omega_{1,2}$ are given by:
\begin{equation}
\omega_{1,2} = \omega({\bf k}_{1,2})
\end{equation}
with $|{\bf k}_{1,2}|^2 = |\ky ~\hat {\bf y} \pm \qx ~\hat{\bf x}|^2 = \ky^2 + \qx^2 $. Note that, in the present case:
\begin{equation}
\omega_1 = \omega_2 = {\rm Re (\Omega)} \simeq 0 
\end{equation}

In the following, we derive the parametric interaction equations.
We obtain, after some algebra, the following system of coupled equations:
\begin{eqnarray}
\frac{\dif \zp}{\dif t} - (\aq \epsilon -\hnu \qx^2 - \hmu) \zp =
\frac{\beta}{4} \hnl \Big[ \psih^* \amp_1  - \psih \amp_2 \Big] \notag\\
- \frac{\hnl}{2} \Big[ \phih^* \pot_1  - \phih \pot_2 \Big],
\label{param11} \\
\left[ \frac{\dif }{ \dif t} + \heta (\qx^2 + \ky^2) \right] \amp_1  +i \freq_1 \amp_1 = \hnl \psih \zp,
\label{param12} \\
\left[ \frac{\dif }{ \dif t} + \heta (\qx^2 + \ky^2) \right] \amp_2 +i \freq_2 \amp_2 =  - \hnl \psih \zp^*,
\label{param13} \\
\left[ \frac{\dif }{ \dif t} + \hnu (\qx^2+\ky^2) \right] \pot_1 + i\freq_1 \pot_1 = \hnl \frac{\ky^2 - \qx^2}{\ky^2 + \qx^2} \phih \zp,
\label{param14} \\
\left[ \frac{\dif }{\dif t} + \hnu (\qx^2+\ky^2)  \right] \pot_2+ i \freq_2 \pot_2 = - \hnl \frac{\ky^2 - \qx^2}{\ky^2 + \qx^2} \phih^* \zp,
\label{param15}
\end{eqnarray}
with the coefficients:
$\hnl = (\hat {\bf z} \times {\bf q}) \cdot {\bf k}_0 = \qx \ky$, and $\freq_{1,2} =0$.
We also included the turbulence drive and neoclassical damping via the term $\aq \epsilon - \hmu$, where $\epsilon = \sum_k |\phi_k^{DW}|^2$ denotes the turbulence energy, $\aq$ is the DW-ZF coupling parameter and $\hmu= \hmu_{neo}$ is the neoclassical friction. In the following, we neglect viscous dissipation for zonal flows since $\hnu \qx^2 \ll \hmu$.

Note that here, since $\freq_1=\freq_2=0$, the two sidebands are directly related via:
\begin{equation}
\psih\amp_2 = - \psih^* \amp_1 \quad {\rm and} \quad \phih \pot_2 = - \phih^*\pot_1
\end{equation}

Hence, only one sideband ($\amp_1, \pot_1$) appears, and the system reduces to: 
\begin{eqnarray}
\frac{\dif \zp}{\dif t} - (\aq \epsilon - \hmu) \zp  =  - \frac{\beta}{2} \hnl \psih^* \amp_1
+ \hnl \phih^* \pot_1,
\label{param21} \\
\left[ \frac{\dif }{\dif t} + \heta (\qx^2 + \ky^2) \right] \amp_1  =  \hnl \psih \zp,
\label{param22} \\
\left[ \frac{\dif }{\dif t} +\hnu (\qx^2+\ky^2) \right]  \pot_1 = \hnl \frac{\ky^2 - \qx^2}{\ky^2 + \qx^2} \phih \zp
\label{param23} 
\end{eqnarray}
The first term on the r.h.s. of the ZF evolution (\ref{param21}) is the direct contribution from the MP-induced nonlinearity  $\{ \psi, \nabla_\perp^2 \psi \}$, in the Maxwell stress-like form whereas the second term on the r.h.s. comes from the indirect $\{ \phi, \nabla_\perp^2 \phi \}$ nonlinearity due to the helical potential $\phih$ associated to the MP, i.e. Eq. (\ref{ident1}).

After some algebra, one can obtain the following 'nonlinear' dispersion relation for zonal flows:
\begin{align}
-i \Omega -   (\aq \epsilon- \hmu) = 
 - \frac{\beta \qx^2 }{2}  \frac{\ky^2 \psih^2  }{-i \Omega + \heta (\qx^2+\ky^2) }, \notag\\
 - \qx^2 \frac{\qx^2 - \ky^2}{\qx^2+\ky^2} \cdot \frac{\ky^2 \phih^2}{- i \Omega + \hnu (\qx^2+\ky^2)}
= 0,
\end{align}
where we replaced $\hnl$ with its expression.

\begin{figure}
\begin{center}
\includegraphics[width=0.5\linewidth]{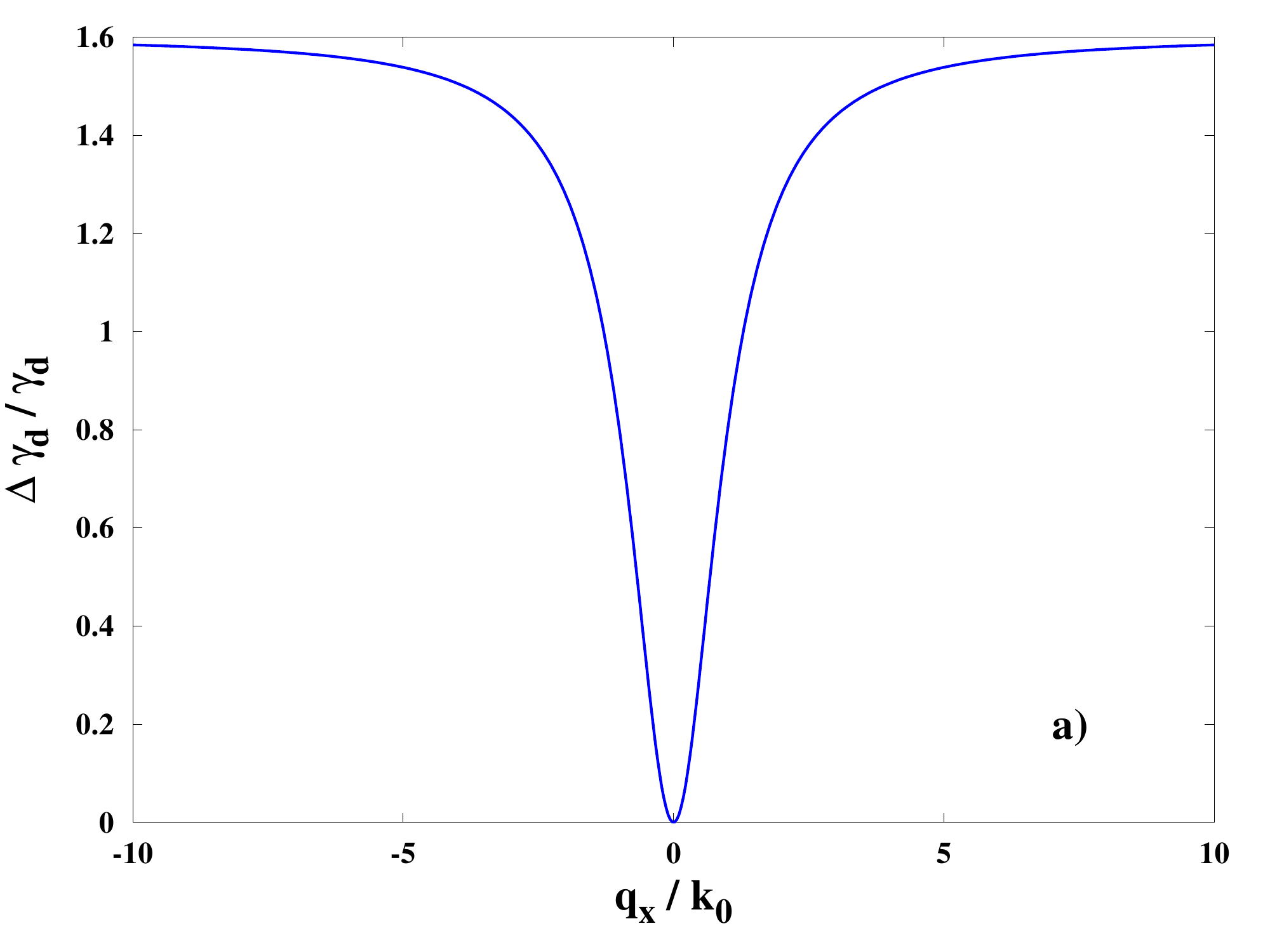}\includegraphics[width=0.5\linewidth]{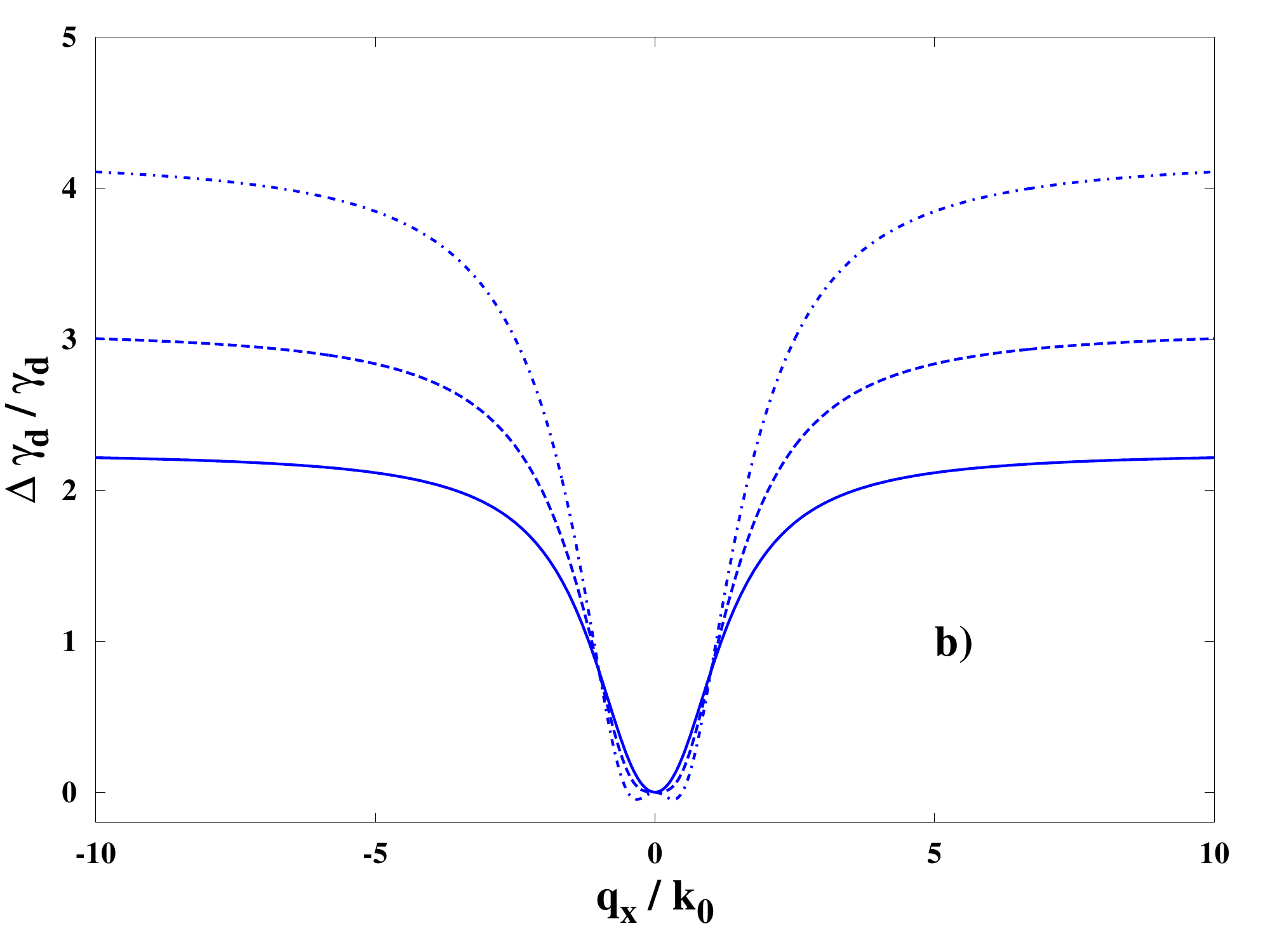}
\caption{Relative change in zonal flow damping $\Delta \gamma_d / \gamma_d$ v.s. ZF radial wavenumber $\qx$, for $B_r^{vac}/B = 10^{-4}$, given by Eq. (\ref{zfdamp-shear}). a) case without mean flow shear $V'=0$ and b) case with mean flow shear $V'=0.1$ (solid), $V'=0.15$ (dash) and $V'=0.2$ (dash-dotted).}
\label{fig-zfdamp}
\end{center}
\end{figure}

In the case without mean sheared flow ($V'=0$, i.e. $\phih=0$, cf. Eq. \ref{ident1}), the associated 'nonlinear' dispersion relation is approximately:
\begin{equation}
\gamma_q - (\aq \epsilon - \hmu) \simeq \frac{ - \beta \qx^2 / 2}{ \heta (\qx^2 + \ky^2)} \ky^2 \psih^2.
\label{zfdamp-noshear}
\end{equation}
with $\gamma_q = {\rm Im} ~ \Omega$ the zonal flow growth-rate.
The enhancement of zonal flow damping is shown v.s. $\qx$ schematically [Fig. \ref{fig-zfdamp}a]. To guide the reader, we can evaluate a typical normalized ZF radial wavenumber as $q_x \rho_s \sim (\rho_s L_n)^{-1/2} \sim 0.2$ or $|q_x / \ky| \sim (a /  nq) / \sqrt{\rho_s L_n} \simeq 20$ for $n=1$ RMPs.

In the limit $|\qx| \gg \ky$, we recover the results of Leconte \& Diamond \cite{LeconteDiamond2012} for the enhancement of ZF damping, in dimensional form:
\begin{equation}
\frac{\Delta \gamma_d}{\gamma_d} \simeq  C_1 \left[ \frac{B_r^{vac}}{B} \right]^2
\label{damping-def}
\end{equation}
with the coefficient $C_1 = c_A^2/(\nu_{ii} \nu_{ei} \lambda_{skin}^2)$ in our notation. Here, $\Delta \gamma_d = \gamma_d - \gamma_d^0$, with $\gamma_d^0$ the reference zonal flow damping without external magnetic perturbation $B_r^{vac} / B=0$.
This reference zonal flow damping is of the order of the ion-ion collision frequency $\nu_{ii}$. The enhancement over this value due to the external perturbation (Eq. \ref{damping-def}) is of the order of $c_A^2/(\nu_{ii} \nu_{ei} \lambda_{skin}^2) \times (B_r^{vac} / B)^2$.
For typical parameters $\nu_{ei} \simeq 5 . 10^5 s^{-1}$, $\nu_{ii} \simeq \nu_{ei} / 40$, $\lambda_{skin} \simeq 10^{-3}$m, and $c_A \simeq 10^6$m/s, this yields:
$\Delta \gamma_d / \gamma_d \simeq 1.6$ for typical external perturbation amplitude $B_r^{vac} / B \sim 10^{-4}$.

In the case with mean sheared flow ($V' \neq 0$, i.e. $\phih \neq 0$), we obtain the following modified 'nonlinear' dispersion relation:
\begin{align}
\gamma_q - (\aq \epsilon - \hmu) \simeq \frac{ - \beta \qx^2 / 2}{ \heta (\qx^2 + \ky^2)} \ky^2 \psih^2 
- \frac{\qx^2 -\ky^2}{\qx^2+\ky^2} \cdot  \frac{ \beta^2 \qx^2 / 4}{ \hnu(\qx^2+\ky^2)} V'^2 \ky^2 \psih^2,
\label{zfdamp-shear}
\end{align}
where we expressed $\phih$ in terms of $\psih$ and $V'$, using the relation (\ref{ident1}).
Eq. (\ref{zfdamp-shear}) is the main result of this Letter.

The effect of mean flow shear on zonal flow damping is shown v.s. $\qx$ schematically [Fig. \ref{fig-zfdamp}b]. Parameters are the same as in [Fig. \ref{fig-zfdamp}a].

In the limit $\ky \ll |\qx|$, the enhancement of zonal flow damping becomes:
\begin{equation}
\frac{\Delta \gamma_d}{\gamma_d} \simeq C_1 \left[ \frac{B_r^{vac}}{B} \right]^2 +C_2 {V'}^2 \left[ \frac{B_r^{vac}}{B} \right]^2
\end{equation}
with $C_1$ given below Eq. (\ref{damping-def}), and the new coefficient $C_2 = L_n^2 / (\rho_s^2 {\nu_i^*}^2) \cdot (q R / L_n)^4$, with $\nu_i^* = \nu_{ii} qR/ v_{th,i}$ the ion collisionality. For typical parameters $R=2$m, $q=3$, $L_n= 5.10^{-2}$m, $\nu_i^* \simeq 0.4$, $V' \simeq 0.1$ and eddy viscosity $\hnu \sim 10 ~m^2.s^{-1}$, this yields $\Delta \gamma_d / \gamma_d \simeq 2.3$, which represents a significant enhancement of zonal flow damping. Morevover, for a flow shear $V' > 1.5$, the relative zonal flow damping becomes negative for zonal flow wavenumbers $\qx < \ky$. Physically, this suggests that the synergy between RMPs and the mean flow shear can excite relativively large-scale zonal flows at wavenumber $\qx < \ky$, while damping short-scale zonal flows, those with larger wavenumbers $\qx \gg \ky$.

\begin{table}
\begin{tabular}{lll}
 & relative zonal flow damping $\Delta \gamma_d / \gamma_d$  \\
\hline
w/o mean flow shear Ref. \cite{LeconteDiamond2012} & $C_1 \left[ \frac{B_r^{vac}}{B} \right]^2$  \\
\hline
with mean flow shear [this work] & $C_1 \left[ \frac{B_r^{vac}}{B} \right]^2 +C_2 {V'}^2 \left[ \frac{B_r^{vac}}{B} \right]^2$
\end{tabular}
\caption{Main scalings of the enhancement of zonal flow damping-rate $\Delta \gamma_d / \gamma_d$ by 3D fields, in the limit $|\qx| \gg \ky$. The coefficients $C_1$, $C_2$ are given in the text.}
\label{tab1}
\end{table}

\section{Discussion and conclusions}
In this work, we used the parametric interaction formalism to derive the zonal flow damping due to external magnetic perturbations. We recovered the results of Leconte \& Diamond \cite{LeconteDiamond2012} in the limit where the poloidal wavenumber $\ky$ of the helical modulation produced by the external field is much smaller than the radial wavenumber of zonal flows $\qx$, i.e. $\ky \ll \qx$. However, our results are more general, as we find that the magnitude of the ZF damping effect shows some dependence on the radial wavenumber of zonal flows, namely short-scale zonal flows are predicted to be more strongly damped by this mechanism than large-scale zonal flows. Moreover, for a sufficiently-large mean flow shear and for large-scale zonal flows $\ky > \qx$, the zonal flow damping becomes negative, i.e. RMPs are predicted to enhance the drive of zonal flows for large mean flow shear, via this mechanism. Collision-free gyrokinetic simulations presented in Ref. \cite{Holod2017} did not observe any effect on zonal flows from external magnetic perturbations. Electron-ion collisions treated in our model may play a role and partially explain this discrepancy. If future improved simulations show a damping, it would be interesting to see if this damping depends on the ZF radial wavenumber.
Due to energy conservation among turbulence/zonal flow system, this additional damping of zonal flows implies a simultaneous increase of turbulence intensity, which can enhance the turbulent transport.

There are limitations to our model (i) We use the vacuum field approximation and thus neglect the plasma response (ii) We do not explicitely treat the spatial resonance aspect of the problem.


In conclusion, we found a new contribution to the zonal flow damping effect due to non-axisymmetric field. This contribution is proportional to the square of the equilibrium $E \times B$ flow shear, and may be important in the pedestal region where $E_r'$ is large. This additional damping of zonal flows implies a simultaneous increase of turbulence intensity, which can enhance the turbulent transport.


\section*{Acknowledgements}
The authors would like to thank Z.X. Wang, M.J. Choi, W.H. Ko and J.M. Kwon for usefull discussions.
This work was supported by R\&D Program through National Fusion Research Institute (NFRI) funded by the Ministry of Science and ICT of the Republic of Korea (NFRI-EN1841-4).

\end{document}